\documentclass[11pt,onecolumn,rmp]{revtex4-1}

\usepackage{natbib}
\usepackage{graphicx}
\usepackage[usenames,dvipsnames,table]{xcolor}
\graphicspath{{Figures/}}
\usepackage{float}

\begin{document}

\title{Taxonomy of Blockchain Technologies. \\
Principles of Identification and Classification}
\author{Paolo Tasca}
\affiliation{Centre for Blockchain Technologies, University College  London, London, United Kindgdom}
\author{Claudio J. Tessone}
\affiliation{URPP Social Networks, University of Zurich, CH-8050 Z\"urich, Switzerland.}

\begin{abstract}
A comparative study across the most widely known blockchain technologies is conducted with a bottom-up approach. Blockchains are disentangled into building blocks. Each building block is then hierarchically classified in main and subcomponents. Then, alternative layouts for the subcomponents are identified and compared between them. Finally, a taxonomy tree summarises the study and provides a navigation tool across different blockchain architectural configurations.


\end{abstract}



\maketitle
\section{Introduction}\label{Intro}
\subsection{Background}\label{Background}
The original combination of a set of existing technologies (distributed ledgers, public-key encryption, merkle tree hashing, consensus protocols) gave origin to the peer-validated decentralised cryptocurrency called Bitcoin, originally introduced by Satoshi Nakamoto in 2008 \cite{Nakamoto2008}. That year was the advent of a new technological milestone: the \textit{blockchain}. Indeed, blockchain has an impact well beyond the specific case of Bitcoin.
 Blockchain allows new forms of distributed software architecture to be developed where networks of untrusted (and sometimes even "corrupted") participants can establish agreements on shared states for decentralised and transactional data in a secure way and without the need of a central point of control or regulatory supervision.  
Blockchain ensures trust among anonymous counterparts in decentralised systems without the need of central supervisor authorities in charge of verifying the correctness of the records in the ledger.
Blockchain has been announced as a disruptive technological innovation, but in fact there is no true technical innovation in Bitcoin and blockchain. All components had already been developed before the Bitcoin paper by Nakamoto in 2009. From a historic perspective, the technology has its roots in Ralph C. Merkle's elaborations, who proposed the Merkle Tree – the use of concatenated hashes in a tree for digital signatures in the 1970s. Hashing has been used since the 1950s for cryptography for information security, digital signatures and message-integrity verification. A decade later, Leslie Lamport proposed using the a hash chain for a secure login. The first crypto currency for electronic cash was described at the dawn of the web in 1990. Further evolutions and refinements of the hash chain concept were introduced in a paper by Neil Haller on the S/KEY application of a hash chain for Unix login, in 1994.
Adam Back proposed hashcash in 2002, but the first eletronic currency based on blockchain with the PoW concept has been proposed by Satoshi Nakamoto's disruptive paper \cite{Nakamoto2008}  While blockchain is still in its emergent technological phase, it is fast evolving with the potential to see applications in  many sectors of our socio-economic systems. According to some statistics summarised by the World Economic Forum  the interest on blockchain expanded globally \cite{WEF_Report}. Almost thirty countries are currently investing in blockchain projects. In the finance sector, 80\% of the banks predict to initiate blockchain-related projects by 2017. Additionally, venture capital investments with a focus on blockchain activities raised to over 1.4 billion USD from 2014 to 2016. Since blockchain digital currencies combine together features of money with those of a payment system \cite{TascaDual}, also central banks started to look into the technology. Currently, over nineteen central banks worldwide do blockchain research and development. Some of them - e.g. the Bank of England - already have commissioned studies on CBDC (Central Bank Digital Currency). From the industry side, over one hundred corporations have joined blockchain working groups or consortia and the number of  patents filled increased to more than three thousand at the moment of writing. These figures show the importance and awareness of blockchain as one of the most promising emerging technologies that, together with  Artificial Intelligence, Internet of Things or nanotechnology will have a pervasive impact on the future of our society.

\subsection{Problem Statement and Research Method}\label{problem statement and research method}
At the moment of writing, we may argue that -- according to the Technology Life Cycle theory --, we are at the beginning of the so called phase of ``fermentation'' which is characterised by technological uncertainty due to the evolution of the blockchain into alternative technological paths. The industry promotes different model designs favouring functional and performance aspects over others in order to meet specific business goals. Currently there are thousands of blockchain projects worldwide under development, some of them run on forks of successful technologies such as Bitcoin or Ethereum, while others propose completely new functionalities and architectures. For this reason, instead of blockchain, in the remainder we refer to \textit{blockchains} or \textit{blockchain technologies} in order to encompass all the possible architectural configurations and, for the sake of simplicity, also the larger family of distributed ledger technologies, i.e., community consensus-based distributed ledgers where the storage of data is not based on chains of blocks. An heterogeneous development combined with a lack of interoperability may endanger a wide and uniform adoption of blockchains in our techno- and socio-economic systems. Moreover, the variation of blockchain designs and their possible configurations represent an hindrance for software architectures and developers. In fact,  without the possibility to resort on a technical reference model, it is difficult to measure and compare the quality and the performance of different blockchains and those of the applications sitting on top of them. To summarise, current variations of blockchain software architectures pose greatest concerns from different perspectives according to heterogeneity:
\begin{enumerate}
\item Heterogeneity is a problem according to the future developments of blockchain technologies, because it will prevent the developments, adoption and stimulation of innovation.
\item Heterogeneity will prevent consistency in drafting laws and policies related to the regulation of blockchain/DLT technologies. 
\item Heterogeneity will increase ambiguity in the application of consumer protection laws and regulations.
\item Heterogeneity will decrease the clarity on how the workforce may be affected by blockchain/DLT technology.
\item Heterogeneity will decrease the clarity in academic research and sharpen concepts that underpin the development of new applications and solutions.
\item Heterogeneity will prevent the development of the specification and use of solutions using blockchain and DLT for ISO, IEC and other SDOs.
\item Heterogeneity will increase the complexity in the understanding of blockchain/DLT for NGOs and how this technology may be applied in the relevant sectors to achieve social and economic goals
\end{enumerate}
The solution to these problems requires the setting up of software reference architectures where standardised structures and respective elements and relations shall provide templates for concrete blockchain architectures.
Standards can emerge naturally because of market adoption (industry driven) or because imposed by institutes and organisations. In the first group we may include initiatives like the Accord Project \footnote{https://www.accordproject.org/}, the ChinaLedger \footnote{http://www.chinaledger.com/} or R3 \footnote{https://www.r3.com/}. In the second group we may refer to the initiative conducted by the International Organization for Standardization (ISO) with the establishment of the technical committee ISO/TC 307 on Blockchain and distributed ledger technologies. Several working groups with different topics to discuss have been settled. In particular, the ISO/TC 307/WG1 working group is engaged with the reference architecture, taxonomy and ontology. Overall, a long-term standardisation of the blockchain reference architecture will benefit every industry.
Thus, a standard for software reference architecture is necessary in order to enable a level playing field where every industry player and community member can design and adopt blockchain-enabled products or services under the same very conditions with possibility of data exchange. As it is for the Internet, several institutes of standardisations (e.g., ETF in cooperation with the W3C, ISO/IEC, ITU) set a body of standards. Internet standards promote interoperability of systems on the Internet by defining  precise protocols, message formats, schemas, and languages. As a result, different hardware and software can seamlessly interact and work together. Applied to World Wide Web (as a layer on the top of the Internet), standards bring interoperability, accessibility and usability of web pages. Similarly,  the adoption of blockchain standards will promote the blossoming and proliferation of interoperable blockchain-enabled applications. Thus, if we envisage a future where blockchains will be one of the pillars of our society's development, it is necessary to begin discussing and identifying standards for blockchain reference architectures. The aim of this study is to highlight the need for standard technical reference models of blockchain architectures. This is timely aligned with the industry sentiment which currently pushes organisations for standardisation to set industry standards. In order to support an appropriate co-regulatory framework for blockchain-related industries, a multi-party approach is necessary as it is for the Internet where both national standards, international standards and a mixture of standards and regulation are in place. In the mid-long term, the lack of standards could bring risks related to privacy, security, governance, interoperability and risk to users and market participants, which can appear as blockchain related cyber crimes.  From a preliminary survey conducted in 2016 by Standards Australia, more than 88\% of respondents indicate the role for standards in supporting the roll out of blockchain technologies \cite{Meguerditchian_Varant}. 
Given the above problem statement the goal of this research is to conduct a review of the blockchain literature. This will be a preparatory work in order to identify and logically group different blockchain (main and sub) components and their layouts.
In order to achieve our goal we propose a blockchain taxonomy. Taxonomy comes from the term "taxon" which means a group of organisms. In our case, taxonomy encompasses the identification, description, nomenclature, and hierarchical classification of blockchain components. This is different from an ontology which would be more focused on the study of the types, properties, and interrelationships of the components and events that characterize a blockchain system.
The methodological approach is composed of the following steps: 
\begin{enumerate}
    \item Analysis across blockchains. A pre condition is the analysis of vocabulary and terms to sort out ambiguities and disagreements. A literature review of the existing technologies is the starting point to limit complexity and organise information in schematic order. To avoid dis-ambiguities the analysis is supported by a merge of common blockchain terminologies developed so far in the literature and grouped in an online database
    \footnote{http//arstweb.clayton.edu/interlex}. This brings together a vocabulary of key blockchain terms to provide readers with a foundation upon which understanding the classification and taxonomy developed in the rest of the analysis. 
The identification of the blockchain components is the crucial part of this analysis.
In order to explore all the possible domains of blockchain components and their topological layout indicating their runtime interrelationships, we conduct a comparative study across different  families of blockchain applications: digital currencies, application stacks, asset registry technologies and asset centric technologies.
See Table \ref{DLT-techs} in Appendix and \cite{tasca2015digital} for more information.
    \item Framework setting. After the comparative study, a hierarchical taxonomy (a tree structure of classifications for a given set of components) has been defined and populated by \textit{main}, \textit{sub} and (when necessary) \textit{sub-sub} components.
    \item Layout categorisation. Finally, for the components in the lowest level of the hierarchical structure, different layouts are introduced and compared. However, as the technology keeps evolving, the layouts are increasing over time. Thus, for the sake of simplicity, we limit our study to two or three main layouts per each \textit{sub} or \textit{sub-sub} component.
\end{enumerate}
\subsection{Results}\label{Results}
The result of the component-level analysis is a universal blockchain taxonomy tree that groups (in a hierarchical structure) the major components, identifies their functional relation and possible design patterns.

In general, it is difficult to evaluate whether a taxonomy or an ontology is good or bad, especially if the domain is a moving target like the blockchain. Taxonomies and ontologies are generally developed to limit complexity and organise information but all serve different purposes and generally evolve over time (see e.g. the evolution of the famous Linnaean taxonomy in biology). Thus, our taxonomy simply aims to contribute to set the foundations for classifying different kinds of blockchain components.  Without claiming to represent the ultimate structure, the proposed taxonomy could be of practical importance in many cases. For example, it can:
\begin{enumerate}
\item support software architectures to explore different system designs and to evaluate and compare different design options;
    \item be propeadeutic to the development of blockchain standards with the aim to increase the adoption at a large scale of blockchain-enabled solutions and services;
    \item enable research into architectural framework for blockchain-based systems in order to boost the adoption of blockhain-enabled systems, their interoperability and compatibility;
\item create gateway models to multiple blockchains and design governance framework;
\item promote blockchain predictability;
\item be used to promote a regulatory framework that provides a mix of both legal and technical rules (i.e., regetech for blockchain-based systems) \cite{marian2015conceptual}.
\end{enumerate}
\section{Background on Blockchain Technologies}\label{DLT-Tech}

Since the Bitcoin inception in 2009, many blockchain software architectures have been deployed to meet different technical, business and legal design options. Given the current complex dynamic of the blockchain architectural development, it would be neither exhaustive nor comprehensive to provide a picture of the existing blockchain technologies developed so far. Therefore, we take a bird-eye view and describe the blockchain by looking at its key driving principles such as data decentralisation, transparency, security, immutability and privacy \cite{Aste-Tasca}.

\textbf{Decentralisation of consensus}. The distributed nature of the network requires untrusted participants to reach a consensus. In blockchain, consensus can be on ``rules'' (that determine e.g., which transactions are allowed and which are not, the amount of bitcoins included in the block reward, the mining difficulty, etc.) or on the history of ``transactions'' (that allows to determine who owns what). The decentralised consensus on transactions governs the update of the ledger by transferring the responsibilities to local nodes which independently verify the transactions and add them to the most cumulative computation throughput (longest chain rule). There is no integration point or central authority required to approve transactions and set rules. No single point of trust and no single point of failure.

\textbf{Transparency}. Records are auditable  by a predefined set of participants, albeit the set can be more or less open. For example, in public blockchains everyone with an Internet connection to the network holds equal rights and ability to access the ledger. The records are thus transparent and traceable. Moreover, participants to the network can exercise their individual (weighted) rights (e.g. measured in CPU computing power) to update the ledger. Participants have also the option to pool together their individual weighted rights. 

\textbf{Security}. 
Blockchain is a shared, tamper-proof replicated ledger where records are irreversible and cannot be forged thanks to one-way cryptographic hash functions.  Although security is a relative concept, we can say that blockchains are relatively secure because users can transfer data only if they posses a private key. Private keys are used to generate a signature for each blockchain transaction a user sends out. This signature is used to confirm that the transaction has come from the user, and also prevents the transaction from being altered by anyone once it has been issued.

\sloppy \textbf{Immutability}. 
Blockchains function under the principle of non-repudiation and irreversibility of records. Blockchains are immutable because once data has been recorded in the ledger, it cannot be secretly altered ex-post without letting the network know it (data is tamper-resistant).
In the blockchain context immutability is preserved thanks to the use of hashes (a type of a mathematical function which turns any type of input data into a fingerprint of fixed size, that data called a hash. If the input data changes even slightly, the hash changes in an unpredictable way) and often of blocks. Each block includes the previous block’s hash as part of its data, creating a chain of blocks. Immutability is relative and relates to how hard the history of transactions is to change. Indeed, it becomes very difficult for an individual or any group of individuals to tamper with the ledger, unless these individuals control the majority of ``voters''.  For public proof-of-work blockchains such as Bitcoin, the immutability is related to the cost of implementing the so-called “51\% attack”. For private blockchains, the block-adding mechanism tends to be a little different, and instead of relying on expensive proof-of-work, the blockchain is only valid and accepted if the blocks are signed by a defined set of participants. This means that, in order to recreate the chain, one would need to know private keys from the other block-adders.
A complete discussion of threats about immutability of the transaction history can be found in \cite{barber2012bitter}. On the other hand, from a governance perspective, this solution is never fully realised. The several examples where the Bitcoin community  had reverted Bitcoin blocks based on community decisions. The division between Ethereum and Ethereum classic, and later between Bitcoin and Bitcoin Cash and Bitcoin Gold are not purely anecdotal evidence: they are strong indicators of the importance that the governing body - even if informal - ends up having on the information eventually stored in the blockchain \cite{walch2017path}.

Other non fundamental properties of blockchain include data automation and data storage capacity. 

\textbf{Automation and smart contracts}. Without the need for human interaction, verification or arbitration, the software is written so that conflicting or double transactions are not permanently written in the blockchain. Any conflict is automatically reconciled and each valid transaction is added only once (no double entries). Moreover, automation regards also the development and deployment of  smart legal contracts (or smart contract codes, see \cite{clack2016smart}) with payoff depending on algorithms which are self-executable, self-enforceable, self-verifiable and self-constraint.

\textbf{Storage}. The storage space available on the blockchain networks can be used for the storage and exchange of arbitrary data structures. The storage of the data can have some size limitations placed to avoid the blockchain bloat problem \cite{BTC_Bloat}. For example, metadata can be used to issue meta-coins: second-layer systems that exploit the portability of the underlying coin used only as “fuel”. Any transaction in the second layer represents a transaction in the underlying network. Alternatively, the storage of additional data can occur ``off-chain'' via a private cloud on the client's infrastructure or on a public (P2P or third-party) storage. Some blockchains like Ethereum allow to store data also as a variable of smart contracts or as a smart contract log event.


\section{Taxonomy of Blockchains}\label{taxonomy}

The diversity of blockchain research and development provides an opportunity for cross-fertilisation of ideas and creativity, but it can also result in fragmentation of the field and duplication of efforts. One solution is to establish standardised architectures to map the field and promote coordinated research and development initiatives. However, in terms of blockchain software architecture design little has been proposed so far \cite{2017-Xu-ICSA}, and the problem of consistently engineering large, complex blockchain systems remains largely unsolved. {We approach this problem by proposing a component-based blockchain taxonomy starting from a coarse--grained connector--component analysis. The taxonomy compartmentalises the blockchain connectors/components and establishes the relationships between them in a hierarchical manner. We adopt a reverse-engineering approach to unbundle the blockchains and divide them into  \textit{main} (coarse-grained) components. Each main component is then split into more (fine-grained) \textit{sub} and \textit{sub-sub} components (where necessary).
 For each of these sub (and/or sub-sub) components, different \textit{layouts} (models) are identified and compared. By deriving the logical relation between (main, sub or sub-sub) components, the study helps to clarify the alternative \textit{modus operandi} of the blockchains and helps to develop the conceptual blockchain design and modelling. 

{ Similarly to other fields like electronics or mechanics \cite{Otto1998}, the software engineering approach used to derive the taxonomy threats blockchains as the result of gluing together prefabricated, well-defined, yet interdependent components.  Although equivalent components provide similar services and functions, they can be of different importance and type, and the  interconnection may work in different ways.
Following this logic, each of the next seven sections will introduce a new blockchain main component and its sub (and eventually sub-sub) components by describing and comparing their layouts.
}

\begin{figure}
\centering
\includegraphics[width=1\textwidth]{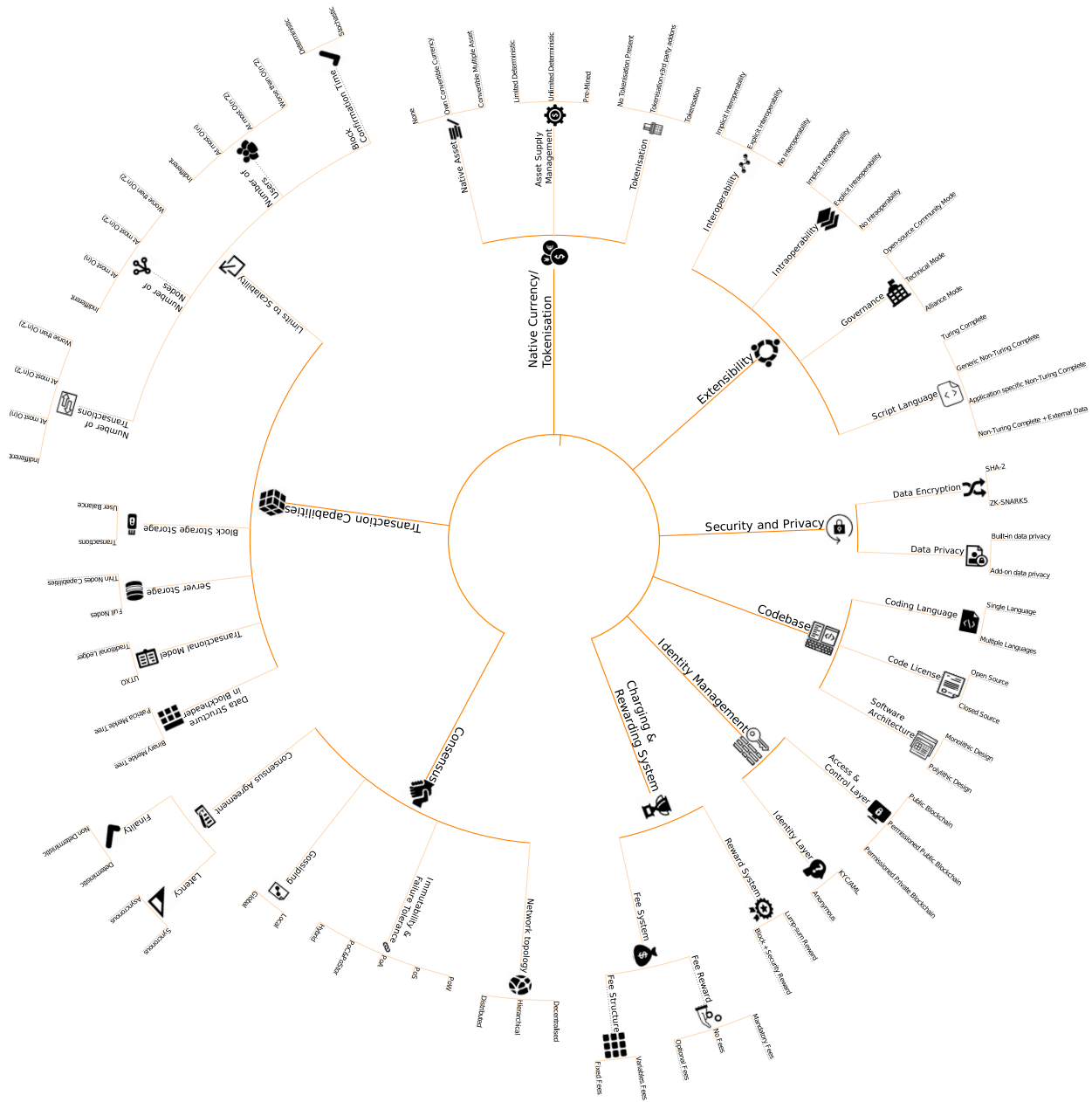}
    \caption{Blockchain Taxonomy Tree: A representation of the taxonomic decomposition of blockchain-based technologies.}
    \label{taxonomy-Matrix}
\end{figure}

\section{Consensus}\label{consensus}
The first identified main component is \textit{Consensus}. It relates to the set of rules and mechanics that allows to maintain and update the ledger and to guarantee the trustworthiness of the records in it, i.e., their reliability, authenticity and accuracy \cite{bonneau2015sok}. \textit{Consensus} varies across different blockchain technologies, every consensus mechanism brings advantages and disadvantages based on different characteristics e.g. speed of transactions, energy efficiency, scalability, censorship-resistence and tamper-proof \cite{Mattila2016}. The set of rules and mechanics compose the framework of the validation process that is necessary to overcome security issues during the validation. Figure \ref{taxonomy-Matrix} illustrates the subcomponents forming the component Consensus: 
\begin{itemize}
       \item[1] Consensus Network Topology
       \item[2] Consensus Immutability and Failure Tolerance
       \item[3] Gossiping
       \item[4] Consensus Agreement
           \begin{itemize}
           \item[4.1] Latency
           \item[4.2] Finality
           \end{itemize}
\end{itemize}
Those subcomponents and sub-subcomponents are to be jointly considered when designing an active network consensus validation process because not only their individual configuration but also their combination determine when and how the overall blockchain agreement is achieved and the ledger updated.

\subsection{Consensus Network Topology}\label{Cons-Netw-Top}\textit{Consensus Network Topology} describes the type of interconnection between the nodes and the type of information flow between them for transaction and/or for the purpose of validation. 
For efficiency reasons, systems have historically been designed in a centralised manner. This centralisation lowers dramatically the costs for system configuration, maintenance, adjustment (and the costs of arbitration in case of conflict) as this work has to be performed only once in a central place. While highly efficient in many situations, this kind of systems induce a single (or very limited set of)  point(s) of failure and suffers of scalability issues. With respect to the network topology, this hierarchical arrangement is still present in most of our techno- and socio-economic systems (one example is the modern electronic payment system). To avoid the single point of failure, these centralised can be extended into hierarchical constructs which exhibit larger scalability and more redundancy, while keeping the communication efficient. 
Alternative to those centralised topologies, decentralised solutions have been proposed. Since the dawn of the Internet, technical systems have evinced a transition towards decentralised arrangements  \cite{Wright2015} where all the nodes are equivalent to any other. For most applications, blockchain based systems - with its federated set of participants - is a clear example of this kind.  Blockchain based systems resort on specific topologies to create the peer network that ultimately determines how the validation process will evolve. 

It is important to mention that \textit{Consensus Network Topology} is linked to the level of (de)centralisation in the validation process but this is not the only determinant. Also other factors like the reward mechanism (see Section \ref{reward}) heavily influence the validation \cite{bonneau2015sok}. As a matter of an example, Bitcoin has a decentralised validation process, which is still be accompanied by ever increasing concentration of computational power devoted to the proof-of-work by network participants. Indeed, during the period 2013-2015, the cumulative market share of the largest ten pools relative to the total market hovered in the 70\% to 80\% range \cite{tasca2015digital}. 

We identify three possible layouts for \textit{Consensus Network Topology}:
\begin{enumerate}
                \item {\bf Decentralised.} There exist implementations that are decentralised. 
                Bitcoin as the pioneer in digital currencies established a distributed P2P network, which enables direct transactions to every node within the network. The validation process within the Bitcoin network is decentralised through miners and full nodes who validate the transactions within the network \cite{Nakamoto2008} connected in a random way as provided by super-nodes. This network illustrates a decentralised \textit{Consensus Network Topology}.  Obviously, this layout is independent from the \textit{Consensus Immutability} layout (Peercoin and NXT also show decentralised network topologies). 
                \item{\bf Hierarchical.} There are other implementations that are not decentralised, and there exists an irregularity of the role  the nodes have. For example, in Ripple \cite{Ripple} the network topology is divided into tracking and validating nodes. The tracking nodes are the gateway for submitting a transaction or executing queries for the ledger, in addition to that the validating nodes have the same functions as tracking nodes but they can also contribute additional sequences to ledger by validation \cite{Ripple}. 
                This kind of solution yields (or can be extrapolated to) a hierarchical network topology. In the community of developers these hierarchical topologies are also referred to as ``Consortium blockchains''.
                \item {\bf Centralised.} In some specific implementations, a central authority may need (or wish) to control what is added to the ledger. An example for this are digital versions of fiat currencies: the so called Central Bank Digital Currencies. This kind of solution yields a third layer, ``Centralised topology'', which is intimately related to private blockchains. It is important to mention that a centralised solution would normally speak of a non-properly working design (or a non-solution) if implemented in terms of a blockchain, as it would have been implemented otherwise in a more transparent manner. Normally, some level of federation and redundancy are key to blockchain systems.
\end{enumerate}

\subsection{Consensus Immutability and Failure Tolerance}\label{Cons-Immut}
 In general, the failure tolerance of a distributed system shall be defined with respect to three interrelated issues: faults (e.g., Byzantine faults), errors and failures. See e.g., \cite{driscoll2003byzantine,castro1999practical} for Byzantine fault tolerance in distributed systems. 
 There are different types of failures and
 generally it is costly to implement a fault tolerant system. Practically, it is not possible to devise an infallible, reliable system . For a literature review and deeper analysis of fault tolerant distributed systems, we refer the reader to \cite{cristian1991understanding} and \cite{fischer1983consensus}.
 A blockchain, as special case of a distributed system, is fault tolerant when it shows the ability to continue functioning. i.e.,  it must grant reliability, validity and security of the information stored in the ledger. 
 Indeed, blockchains represent a  decentralised solution to the problem of information storage which require no central database but many duplicates such that each server holds a \textit{replica} of the ledger. Any new record is costly (often measured in terms of computational power) to be added to the ledger, but cheap to be verified by peers.
 Therefore, a blockchain system is in the need of an efficient consensus mechanism to ensure that every node has its original version of the full transaction history which is kept consistent with the other peers over time. In this vein, the immutability of achieved consensus differs with respect to the  resources required to keep  large network security. 
 In the past years, the evolution of blockchain technologies  has been accompained with the development of different mechanisms that help to keep reliable, valid and secure the information contained in the ledger. 
All together, the mechanisms for \textit{Consensus Immutability}  together with the subcomponents  of  \textit{Consensus Agreement} determine the failure tolerance of the blockchains.
As of this writing, we identify six main layouts for \textit{Consensus Immutability and Failure Tolerance}:
\begin{enumerate}
                \item {\bf Proof-of-Work}. The most widely used cryptocurrency, Bitcoin, uses Proof-of-Work (PoW) to ensure  the immutability of the transaction records. In this setup, computing devices, usually called \textit{miners}, connected to a peer-to-peer network perform the task of validating the transactions proposed for addition to the complete record of existing - valid - transactions. 
                The generation of a block that can be appended to the blockchain - rendering in this way valid all transactions there included - requires {finding the solution of inverting a cryptographic function, which can only be done by brute force. In PoW, the probability that a miner mines a new block depends on the ratio between the computational power he devotes to this task and the total instantaneous computational power by all miners connected to the network. 
                Specifically, miners must find a solution to an one-way hash function by computing new hash values based on the combination of the previous hash values contained in the message, the new transactions in the block they create and a nonce. The solution is such that the new hash value will start with a given number of zeros $\leq$ target.  
                As for this writing, the mining process needs several requirements to be successful \cite{BitFuryGroup}. These include specialised hardware which is needed to perform the computational tasks and ever increasing amounts of electricity to power the hardware.
                These computations are run by dedicated machines (ASICs) which are very expensive and are resource-intensive as contribute to a large electricity footprint for cryptocurrency miners \cite{o2014bitcoin}.  
                Due to this scheme, in the last years miners agglomerate around mining pools \cite{Lewenberg:2015}. 
                 Therefore, a clear drawback of the PoW mechanism is the inherent inefficiency from the resource point of view, and the large-scale investments needed, which has led to long-term centralisation of the mining power. In late 2017, every second almost five quadrillion SHA256 computations were performed in the Bitcoin mining process. Regretfully, these computations do not have any practical or scientific relevance apart from ensuring that the process of block creation is costly, but others' blocks validity are simple to verify for peers. 
                 }
               Interestingly, when adversaries coordinate, it is sufficient that they hold only the 25\% of the total computing power to mount an attack \cite{eyal2014majority}.
               {In this layout, there exists the risk of monopoly mining, induced by large coordination of miners in a single mining pool, which continuously increases the expected payoff of others if they join said mining pool. In this hypothetic situation, said maining pool can censor specific transactions and dictate what transactions are accepted and which ones not.  
               }
               In contrast, BFT consensus mechanisms tolerate at most $n/3$ corrupted nodes in the asynchronous communication protocol and even higher levels in the synchronicity case.
               Electricity consumption can be estimated around 0.1 to 1 W/GH corresponding to around 1GW of electricity consumed every second. Therefore, other developers within the area of blockchain technologies continuously attempt to develop novel mechanisms to achieve an equivalent goal. {It is worth mentioning that some cryptocurrencies (e.g. Primecoin) have tried to make the PoW a task that serves a useful aim (in that case, searching for long chains of prime numbers, or Cuningham series).}
                \item {\bf Proof-of-Stake}. PoS links the block generation to the proof of ownership of a certain amount of digital assets (e.g., digital currencies) linked to the blockchain. {The probability that a \textit{prover} is selected to verify the next block is larger the larger is the share of assets this prover has within the system. The underlying assumption is that users with a large share of the system wealth are more likely to provide trustworthy information with respect of the verification process, and are therefore  to be considered trusted validator \cite{Mattila2016}. } Two alternative PoS methods have been devised. {The first one is based on randomised block selection (used in e.g., NXT and BlackCoin); it uses a calculation searching  for the lowest hash together with the stake size; it is therefore somewhat deterministic and each node can independently determine the likelihood of being selected in a future round. An alternative scheme is the \textit{coin-age}-based selection (used by e.g., Peercoin, being actually the first one to be implemented in real world) which combines randomisation with coin-age (a number derived from multipliying the amount of the assets held by the prover and the length of time it has been helding them).} 
  Although PoS has the chance to solve two issues with PoW (risk of monopoly mining and resources wasted in the mining process), it is affected by the ``nothing at stake'' issue. Because there is little cost in working on several chains (unlike in PoW), one could abuse by voting for multiple blockchain-histories which would prevent the consensus from ever resolving (double spending). This problem can be addressed by Delegated Proof-of-Stake (DPoS), a generic term describing an evolution of the basic PoS consensus (utilised in, e.g., BitShares, Casper by Ethereum, Tendermint) where blocks are forged by a predetermined users delegated by the user who has the actual stake. These forgers are rewarded for their duty and are punished for malicious behaviour (such as participation in double-spending attacks). This principle of pre-authorised forgers is generalised by the Proof-of-Authority mechanism. 
                \item{\bf Proof-of-Authority}.  In this case, participants are not
asked to solve arbitrarily difficult mathematical problems like in
PoW, but instead they are asked to use a hard-configured set of
``authorities'' empowered to collaborate ``trustlessly''. Namely, some nodes
 are exclusively allowed to create new blocks and secure the
blockchain. Typically, Proof-of-Authority (PoA) mechanism
fits well for consortium private networks where
some preselected real entities (i.e., the \textit{authorities}) are  allowed to
control the content that is added to the public registry. Those nodes will receive a
set of private keys that will be used to ``sign" the new blocks, acting as \textit{trusted signers}. 
Thus, every block (or header) that a client
sees can be matched against the list of trusted signers. The
challenges brought by PoA are related to: control of mining
frequency, distribution of mining load (and opportunity) between
the various signers and; maintenance of the the list of signers
such to be robust from malicious attacks even in presence of
dynamic mutation of the trusted signers. 
\item{\bf Proof-of-Capacity/Proof-of-Space and Proof-of-Storage}. 
PoC or PoSpace and PoStorage 
are implementations of the popular idea of ``space as resource''. Here the focus is not on the CPU cycles but on the amount of actual memory (non-volatile) space
the prover must employ to compute the proof. Nodes are asked to allocate significant volume of their hard drive space to mining instead of using CPU-bound space as in PoW. Miners are incentivised to devote hard-drive capacity as those who dedicate more disk space have a proportionally higher expectation of successfully mining a block and reaping the reward. The PoC makes use of hash trees to efficiently allow verification of a challenge
without storing the tree. These schemes are more fair and green than PoW. The reason mainly comes from the lower variance of memory access times between machines and the lower energy cost achieved through the reduced number of computations required. Several practical implementations adopt the PoC consensus algorithm like Permacoin, SpaceMint and Burstcoin, just to cite a few.
PoC consists of an initialisation and subsequent execution between a prover $P$ and a verifier $V$ \cite{dziembowski2015proofs}.  Rather than $P$ proving to $V$ that some amount of work has been completed, $P$ proves to $V$ that she has allocated some number of bytes of storage.  After the initialisation phase, $P$ is supposed to store some data $F$ of size $N$.  Instead, $V$ only holds some small piece of information. At any later time point $V$ can initialise a proof execution phase, and at the end $V$ outputs reject or accept. The PoC is in general defined by three quantities: ($N_0$, $N_1$, $T$); then, the miner shows that she either: 1) had access to at least $N_0$ storage between the initialisation and execution phases and at least $N_1$ space during the execution phase; or 2) used more than $T$ time during the execution phase.  Solutions to the ``Mining multiple chains'',  and ``grinding blocks'' problems of PoC algorithms have been proposed by \cite{park2015spacecoin} among the others. The Proof-of-Storage (PoS) mechanism is similar to PoC but the designated space in it is used by all participants as common cloud storage \cite{PoC}.
 \item {\bf Proof-of-Burn}. In Proof-of-Burn (PoB) miners must prove that they burned some digital assets. They do so by sending them (e.g., digital currencies) to a verifiable unspendable address belonging to them.  Similarly to the PoS, also the PoB logic is to minimise the waste of resources generated by PoW. However, at the current stage, all PoB mechanisms function by burning PoW-mined digital currencies. This is therefore an expensive activity as the digital currencies once required to work as ``fuel'' in a PoB system cannot be recovered \cite{bonneau2015sok}. PoB can be used also to bootstrap a token off of another (see e.g., Counterparty or Mastercoin).

 \item {\bf Hybrid}. The more advances hybrid consensus immutability and failure tolerance methods are ``PoB and PoS'' where Proof-of-Burn blocks act as checkpoints and ``PoW and PoS'' where PoW blocks act as checkpoints containing no transactions, but anchor both to each other and to the PoS chain. Peercoin uses PoW/PoS consensus. To solve the ``nothing at stake'' issue, Peercoin uses centrally broadcast checkpoints (signed under the developer's private key) according to which no blockchain reorganisation is allowed deeper than the last known checkpoints. Here the problem is that the developer becomes the central authority controlling the blockchain.
\end{enumerate}

\subsection{Gossiping}\label{Gossiping}
Blockchains are also decentralised, redundant storage systems. This redundancy makes it very difficult to hijack the information stored in them. 
How this information travels through the network of computers is a characteristic that varies from one blockchain system to another. 
Given the lack of a central routing authority (like it would exist for example in traditional electronic payment systems) nodes must transmit the information they possess -- in general new blocks, but it may be also the full blockchain to new nodes that enter  the network -- to peers they know are participating of the system. 
To this aim, nodes possess a list of peer nodes. 
Whenever a new block is added to the local blockchain of a node, the later passes the block to others in its peer list by \textit{Gossiping}.

We identify two possible layouts for \textit{Gossiping}:

\begin{enumerate}
                \item {\bf Local.} \textit{Gossiping} occurs first in a local manner (through a local validation process) until consensus is reached. This is also called ``federated consensus" used e.g., in Ripple \cite{Ripple} in which nodes can share transaction records to another node and reach consensus without directly knowing all the nodes in the network. Therefore most information travels ``locally" -- in terms of the P2P network -- such that a  consensus is reached at this initial level. Only then, the information is sent throughout all the other nodes. In this layout, the \textit{Gossiping} can be termed ``local''.
                \item {\bf Global.} In most implementations -- Bitcoin, Ethereum, etc. -- \textit{Gossiping} occurs to a list of peers that have been selected by what in the Bitcoin network are called fallback nodes. {These fallback nodes maintain a list of all peers in the network. Upon connection of a new node, the submit a randomly chosen list of peers to the entrant one. The logical network topology is intended to be largely unstructured, similar to the Erdos-Renyi network \cite{barabasi2016network}. Such topology lacks a concept of vicinity or local neighbourhood, and therefore the \textit{Gossiping} process can be termed \textit{global}.       }
\end{enumerate}

\subsection{Consensus Agreement}\label{Cons-Agreement}
The \textit{consensus agreement} defines the set of rules under which { records (like sets of transactions or any other atomic piece of information) are independently updated by the nodes of a distributed systems. This is important to understand how a distributed system is able to handle the so-called Byzantine failures, i.e., how the system composed of $n$ nodes can achieve consensus on storing verified, trustworthy,  information even in the presence of $f$ malicious nodes or in presence of malicious participants launching sybil attacks \cite{10.1007/3-540-45748-8_24}. 
In this regard, it is very important to understand how the nodes communicate between them.}

\subsubsection{Latency}\label{Lat-cy}
\textit{Latency} is a sub-subcomponent  which describes the rule of message propagation in the networks.
\begin{enumerate}
                \item {\bf Synchronous Communication.} Systems which set 
                 upper bounds on ``process speed interval''  and ``communication delay'' such that every message arrives within a certain known, predefined, time-interval ($\Delta$) are called \textit{synchronous}.
 This does not preclude the possibility of having message delays due to exogenous network latency, but the delay is bounded and any message that takes longer than $\Delta$ is discarded. Lax synchronicity assumptions apply also to the Bitcoin blockchain. For example, a block is rejected if contains a timestamp: 1) lower than (or equal to) the  median timestamp of the previous eleven blocks; and 2) greater than (or equal to) 
 the ``network-adjusted time'' plus 2 hours.  Another example of blockchain adopting \textit{synchronous communication} is Ripple through the use of clocks. Specifically, Ripple's ``LastLedgerSequence'' parameter asserts that a transaction is either validated or rejected within a matter of seconds.
                \item {\bf Asynchronous Communication.} Systems which do not set any bound on ``process speed interval'' and ``communication delay''
                such that every message/packet can take an indefinite time to arrive are called \textit{asynchronous}. Although this type of communication protocols brings some advantages (e.g., calls/requests do not need to be addressed to active nodes and nodes do not need to be available when a new information is sent to them by peers), its main disadvantage is that response times are unpredictable and it is harder to design applications based on them. Synereo is an example of blockchain using the asynchronous communication  protocol. 
\end{enumerate}

\subsubsection{Finality}\label{agree-m}
{\textit{Finality} describes whether information intended to be stored in a blockchain (or, as a matter of fact, in any system) can be safely considered \textit{perpetually} stored once the recording is performed. For a distributed system like blockchain-based ones this is very challenging to achieve, and it is certainly not one of the underlying design principles. In a system where the new blocks diffuse through gossiping, and because of rules such as the precedence of the longest chains, even if consensus is achieved globally, \textit{a priori} nothing prevents a set of new nodes entering the system and overriding the previous consensus by offering a longer versions of the hostory. } 
We identify two possible layouts for \textit{Finality}:
\begin{enumerate}
                \item {\bf Non-Deterministic}. In this case, \textit{consensus agreement} ``eventually settles''.  Non-deterministic are randomised or inherently probabilistic consensus (also called \textit{stabilising} consensus) in which the probability to disagree decreases over time. For example in the Bitcoin blockchain 
 the block frequency is adjusted (with respect to the block-mining rate and indirectly to the computational power of the nodes)  to minimise the probability of forks. 
Moreover, the propagation of blocks through the network has characteristic delays \cite{decker2013information} and
 { even in presence of only honest nodes the fork probability cannot be ruled out simply because different nodes may find competing blocks of the same height before the one found first reaches the complete network . This cannot be prevented even if there is in place a concurrency control mechanism, which which attempts to correct results for simultaneous operations. Therefore, overall, the protocol is non deterministic. Thus, even though the widespread heuristic ``wait until 6 confirmed blocks are appended to the chain" reduces the likelihood that a transaction is overridden afterwards, it does not eliminate completely the probability of a previously validated block to be pruned and removed from the blockchain in the future.}
                \item {\bf Deterministic}. In this case, \textit{Consensus Agreement} converges with certainty and transactions are immediately confirmed/rejected in/from the blockchain. This property turns to be very useful for smart contracts where, using state-machine replication, consistent execution of the contracts can be achieved across multiple nodes.
               { All the blockchains based on Lamport Byzantine Fault Tolerance \cite{lamport1982byzantine} achieve deterministic consensus. A prime example of an implementation featuring deterministic finality is Stellar. Another case of deterministic is for private blockchains where new blocks follow a predefined set of rules.}
\end{enumerate}

\section{Transaction Capabilities}\label{Trans-Cap}
The second main component, \textit{Transaction Capabilities}, is important to illustrate  scalability of transactions and  usability in possible applications and platforms. One of the major challenges for the blockchain technology is to increase the  transaction throughput to compete with other solutions already available in the market{ (e.g. centralised payment systems, like credit cards).} In order to achieve these improvements,  quantitative parameters (e.g. data storage in block header, TPS (transactions per second)), need to  be redesigned to realise such improvements. Figure \ref{taxonomy-Matrix} illustrates the subcomponents forming the component Transaction Capabilities: 
\begin{itemize}
       \item[1] Data Structure in the Blockheader
       \item[2] Transaction Model
       \item[3] Server Storage
       \item[4] Block Storage
       \item[5] Limits to Scalability
 \begin{itemize}
           \item[5.1] Transactions
           \item[5.2] Users
           \item[5.3] Nodes
           \item[5.4] Confirmation Time
           \end{itemize}
\end{itemize}

\sloppy \subsection{Data Structure in the Blockheader}\label{data-structure}
The data stored in the block header has different functions. On the one hand, it includes the transaction hashes  for validation purposes; on the other, it contains additional information for different application layers or blockchain technology platforms. The \textit{data structure in the blockheader} describes the capabilities of the system to store transaction information. The  original application of Merkle proof was implemented in Bitcoin, as described  in  \cite{Nakamoto2008}.
We identify two possible layouts for \textit Data Structure in the blockheader:
\begin{enumerate}
\item {\bf Binary Merkle Tree.} Bitcoin uses the Binary Merkle tree \cite{10.1007/3-540-48184-2_32} within the block header to store the transactions. The information in the block header in the Merkle tree structure contains a hash of the previous header, timestamp, mining difficulty value, proof of work nonce and root hash for the Merkle tree containing the transactions for that block, which are used for the verification process to scale up the transactions speed. By convention, the longest chain (since the so-called Genesis block) is considered to be the current status of the blockchain. 
\item {\bf Patricia Merkle Tree.} One the one hand, \textit{Patricia Merkle Tree} (Practical Algorithm To Retrieve Information Coded In Alphanumeric \cite{morrison1968patricia}) allows activities like inserting, editing or deleting information referring to the balance and nonce of accounts, which enables faster and more flexible validation of transactions than the one \textit{merkle tree model} \cite{wood2014ethereum}. However, with respect to the applications, it has the important advantage of allowing for verification of specific branches of the tree. Ethereum \cite{Ethereum} uses the Patricia Merkle Tree within the block header to store more information than what is possible in the Binary Merkle Tree. Those contain transactions, receipts (essentially, pieces of data showing the effect of each transaction) and state \cite{ethDec}. 
Importantly, this technology allows even blocks outside the longest chain to contribute to the validation process, building a confirmation system that is less centralised. This is the so called Ghost rule, a variant of which is implemented also in the Ethereum blockchain \cite{ethDec}.
\end{enumerate}


 \subsection{Transaction Model}\label{transc-model}
{The transaction model can be imagined as an accounting ledger which tracks the inputs and outputs of each transaction. The \textit{transaction model} describes how the nodes connected to the P2P network store and update the user information in the distributed ledger.
The challenge of the \textit {transaction model} is to prevent data that ought not to be trusted by the parties connected to the system - e.g.~those originated in  behaviour, like double spending - to enter into the ledger 
}
As of this writing, it is possible to identify two possible layouts for \textit{Transaction Model} in the widely used blockchain-based systems:
\begin{enumerate}
                \item {\bf The Unspent Transaction Output (UTXO).} \textit{UTXO} model includes a refractory number of blocks during  which network participants are prevented of using the transaction output in new transactions. In this way, it prevents miners from spending transactions fees and block rewards before stable validation status of the block chain. This measure prevents the \textit{forking problem} of blockchains \cite{BLCHGuide}. This transactions mechanism is available in blockchain technologies like Bitcoin. 
                \item {\bf Traditional Ledger.} {In comparison to the \textit{UTXO model}, different implementations of blockchain systems - like \textit{Stellar} and \textit{Ripple} use a more traditional ledger model to record the transactions recorded in the system}. In particular, Stellar lists every single transaction in the \textit{Stellar distributed ledger history}. Also, \textit{Ripple} uses the traditional ledger transaction model to register increments/decrements of balance and clear all account balances. In \textit{Ethereum} some transactions are used to execute actions in smart contracts defined in specific atomic records in the blockchain. Those transactions can be seen as order executions of stakeholders which perform the actions out of said smart contracts.
\end{enumerate}

\subsection{Server Storage}\label{server-store}
At the core of blockchain-based systems underlies their decentralised nature. This requires that nodes connected to the peer-to-peer network are indistiguishable from each other. 
This concept, however, cannot be fully expressed  when the storage needs, computing power or bandwidth constraints of the network nodes do not permit this feature to be fully realised. 
In these scenarios, different nodes have access to different layers of information, those which do not store the information fully are ``thin clients'' connected to the peer-to-peer network \cite{xu2018blockchain}.
We identify two possible layouts for \textit{Server Storage}:
\begin{enumerate}
    \item {\bf Full Nodes.}  All nodes connected to the network, and which are part of the validation process, are of the same kind. This is a genuinely peer-to-peer network where all the nodes are equivalent in terms of information contained. This property creates a large information redundancy, which makes the system more resilient to attacks or malfunctioning.
    \item {\bf Thin Nodes Capabilities.} In this setup, some nodes connected to the network contain only a selected subset of all the information contained in the blockchain. This creates more scalable systems (in terms of number of nodes connected to the network and the concomitant network traffic and storage needs), but may deteriorate the resiliency as only  a fraction of the nodes contain the complete blockchain information.
\end{enumerate}

\subsection{Block Storage}\label{Block-store}
Which information is stored in the blockchain determines the scalability of the system across some dimensions. More crucially, it also allows to understand how concomitant information from users are abstracted within the system.
We identify two possible layouts for \textit{Block Storage}:
\begin{enumerate}
    \item {\bf Transactions.} In systems like Bitcoin, only the transactions are stored. They contain both, a set of inputs and outputs that help to identify emitter(s) and receiver(s) of a specific transaction. {This kind of approach is preserved in more exotic applications of blockchain-like technologies, like IOTA, which relies on the storage of an directed-acyclic-graph to store every single transaction. This kind of approach works not only for cryptocurrencies applications, but it is underlying all transfer-of-property-like applications.}
    \item {\bf User balance.} In systems like Ripple, the decentralised storage also contains information about the user balance in the specific assets. {This approach may limit the storage needs of the system, but at the same time reduces accountability and the possibility to roll back transactions.}
\end{enumerate}

\subsection{Limits to Scalability}\label{Limit-scale}
The decentralised nature of blockchain systems and the concomitant redundancy in the storage impose different kinds of limits to the way in which a specific implementation scales when the \textit{system size}.
System size is used here in a broad sense: It {may refer to} to the number of nodes connected to the network, the number of users of the service, the set of network connections and/or amount of network traffic, the number of transactions, etc.
It is worth remarking that these ingredients are intertwined in the real world \cite{Tessone2017b}, and - upon usage and continuous development - the limiting factor of a particular blockchain system may vary over time. {In a rapidly evolving technology such as blockchain is nowadays, these limits are often changed by the development teams behind some of these systems. An example is the implementation (or not) of SegWit2x and Lightning (as technology for micropayment channels) \cite{decker2015fast} as ways to alleviate the limitation in the number of transactions that Bitcoin can process with respect to its current implementation.  }

{The importance of this component is due to its influence on the final scaling of the system}. Scaling is a property that specifies how the growth will influence its overall performance. As an example, how the total network traffic induced by unverified transactions grow with the number of network nodes. If every node has a small - limited - number of connections, then the total network traffic will scale linearly on the number of nodes. In mathematical terms it will be $\mathcal{O}(N)$. 
However, if every node is connected to each other then the traffic will be $\mathcal{O}(N^2)$, i.e.~it will grow quadratically on the number of nodes. Therefore, if - for a given implementation - network traffic is the most crucial limiting factor, different logical topologies will have different   scalability. Acknowledging that a categorical definition is a crude simplification, we will focus here on the most limiting element for each system.
We identify four possible layouts as \textit{Limits to Scalability}:

\sloppy \begin{enumerate}
    \item {\bf Limit by number of transactions.} We start from the most common real-world example. Bitcoin has a limitation in the number of transactions it can process in every block, because of the hard-coded limit to the block size in bytes. Given that new blocks appear (on average) every ten minutes, this means that the number of transactions that can be included in a given time window is limited. Therefore the layout ``Number of Transactions'' refers - regardless of the information stored in the blockchain\cite{Eyal2015} - to the specific implementations, where the number of operations that can be included in the blockchain is severely limited by design.
    \item {\bf Limit by number of users.} Bitcoin only stores transactions in its public ledger. This is different from other related technologies. Ripple, on the other hand, stores not only transactions, but also the state of the Ripple accounts. Therefore, in scenarios like this, it is the number of users of the system that limits its scalability. A similar problem occurs in Ethereum where the system will be constrained by the number of DAOs, individuals, etc. that it will contain as these are the actors that generate activity in the system. Therefore, the term ``Number of Users'' for this layout is a broad reference to the number of objects of which states stored. Needless to say, this layout is somehow related to the previous, the number of transactions will depend on the number of users. However, one limiting factor can still appear irrespective of the other.
    \item {\bf Limit by number of nodes.} The number of nodes connected to the network, acting as verifiers for the information that is stored on the blockchain, presupposes a limiting factor because of the mechanism of information diffusion adopted. \textit{Gossiping} is a process that requires larger times in decentralised networks to propagate into a consensus state \cite{Tessone2017}, and may even reach a point - where the relative time taken by network traffic is very long - where consensus cannot longer be reached and the blockchain naturally forks. Therefore this process naturally limits the applicability of fully decentralised solutions. 
    \item {\textit{Possible values.}} These three layouts can have three different values, regarding on how detrimental is a specific layout to the overall performance of the system.  The possible values that each layout can have are divided into four: (i) \textbf{Indifferent}, (ii) {\bf At most linear},  (iii) \textbf{At most quadratic},  (iv) {\bf Worse than quadratic}. The first value is assigned \cite{catalini2874598some} when the relevant global characteristics of a  system is independent of the number of specific class; the other three, express three categorical values that are assigned to the dependency of the number of elements in said class. For example, the number of users is largely irrelevant to the performance of the Bitcoin network (because this number is never translated into any property of the network). However, the number of transactions increases linearly a penalty on the local network traffic.
    \item {\bf Confirmation time.}  The time it takes a specific action to be  confirmed ultimately depends on the time it takes for it to be added to the blockchain, and to be validated to further blocks later appended to it. Different approaches can be taken to this process: \textbf{deterministic} addition of new blocks at regular intervals (taken by Peercoin) and \textbf{stochastic} addition like in Bitcoin, where the process of mining induces an Exponential distribution of inter-block discovery time. 
\end{enumerate}

\section{Native Currency/Tokenisation}\label{native-currency-tok}

{
So far, cryptocurrencies and other transfer of property records are the most common usage of the blockchain technology. In cryptocurrencies, system participants who contribute to the verification process - if selected by some rule to issue a new block into the blockchain - are awarded the possibility to issue a transaction without issuer (so called ``coinbase'') to themselves. On the one hand, this is a customary way of introducing new assets into the system. On the other, it introduces an incentive for users to participate of the verification process which leads to an increased trustworthiness on the system.}

{The aforementioned incentive scheme \cite{Sompolinsky2018} is to be provided in a token, whose value is assigned \cite{catalini2874598some} precisely because of the cost associated with its production \cite{garcia2014digital}. Initially, solutions like Bitcoin have created its own (and single) asset class (\textit{the bitcoin}) that can be transacted within the system. This particular solution is not the only one possible with the primary example of Ethereum, where beyond the natural Ether native token, via smart contracts arbitrary new tokens can be created and their property exchanged.
Further, the  native currency possibilities present for example in Ripple \cite{tsukerman2015block} and tokenisation enable different use cases of the blockchain technology like asset-transfers via tokens, exchanges, etc.
All this is just the beginning of cryptoeconomics: it is of uttermost importance  how these assets are supplied into the system, because this affects the way users are incentivised to participate in the validation process. } Figure \ref{taxonomy-Matrix} illustrates the subcomponents forming the component Native Currency/Tokenisation:
\begin{itemize}
       \item[1] Native Asset
       \item[2] Tokenisation
       \item[3] Asset Supply Management
\end{itemize}

\subsection{Native Asset}\label{nat-cur}
Some systems implemented using blockchain technologies have underlying a native asset (which are normally called \textit{cryptocurrency})  which is a digital  token whose owners assign a value and  allow to run the daily activities on the platforms or communities.
Whether these cryptocurrenties ought to be considered fiat or commodity currencies \cite{grinberg2012bitcoin,selgin2015synthetic,Luther2018}, and whether they may eventually be massively adopted replacing traditional ones \cite{Luther2016}.
We identify three possible layouts for \textit{Native asset}:
\begin{enumerate}
    \item {\bf None.} Private blockchain implementations do not require a native asset within to incentivise participation. In these cases, there is no native asset incorporated into the system
    \item {\bf Own Cryptocurrency.} Most implementations of cryptocurrencies only deal with transfer of property of its own tokens within the system.  
    Bitcoin or Litecoin are examples of technologies with single asset compatibility \cite{buterin2014next}. These technologies are limited to their own underlying digital currency, but it can also have off-chain solutions to interoperate with other currencies to execute transactions or to enrol into smart contracts. Further, solutions like coloured-coins \cite{rosenfeld2012overview}.
    \item{\bf Convertible Multiple Assets.} Other technologies like Counterparty, Ardor o do have their own underlying currencies or tokens to execute tasks. However, these technologies also enable the possibility of exchange of assets expressed in others outside those native to the platform. This approach of  multiple, convertible, currencies has the advantage of allowing for exchange markets be directly reflected into the system.
\end{enumerate}



    


\sloppy \subsection{Tokenisation}\label{tokens}
A token acts as a digital bearer bond, whose ownership is determined by the data embedded in the blockchain. Ownership of the tokens is  transferable between holders using other transactions with associated ``transfer'' metadata. This does not require the approval of any other authority. The possibility of tokenisation\cite{catalini2874598some} enables a range of possible use cases for the blockchain technologies  outside the purely financial world\cite{tsukerman2015block,rohr2017blockchain,adhami2017businesses,conley2017blockchain}.

We identify three possible layouts for \textit{Tokenisation}:
\begin{enumerate}
  \item {\bf No tokenisation present.} Without third-party technologies , Bitcoin does not have implemented technologies that enable tokenisation. 
  \item {\bf Tokenisation through third-party addons.} Bitcoin plus Colour-Coin \cite{rosenfeld2012overview} enables the existence of tokenised transactions in the Bitcoin blockchain. Such solution is based on the cryptographic nature of Bitcoin addresses and the script language.  
  \item {\bf Tokenisation.} The tokenisation possibilities together with the extensions of metadata are available in several implementations and constitute the backbone of blockchain-based property registries.  The most paradigmatic example is Ethereum, where the creation of a new Token is produced by means of the creation of a smart contract. Thanks to this flexibility, and extreme extension possibility of such platform, the conditions for creation of new tokens is countless.
\end{enumerate}

\quad

\sloppy \subsection{Asset Supply Management}\label{monetary-supply}
The process of the digital asset (usually referred to as \textit{cryptocurrency}) creation varies across different blockchain technologies. 
Each approach has taken different economic frameworks in most cases fixing a specific monetary policy the future of a particular system. 
This is also a pillar of the incentive scheme that users have to participate (or not) in the validation process \cite{Tessone2017}.

We identify three possible layouts for \textit{Asset Supply Management}:
\begin{enumerate}
    \item {\bf Limited - Deterministic.} The most replicated system in the world of blockchain is the limited supply as introduced in Bitcoin. Not only the supply grows sub-linearly over long periods of time (in contrast to what occurs in normal fiat currencies), but it is designed to have a well defined limit. It is important that, while this incentivises users to adopt the technology and contribute to the process of verification - for which they get a retribution -, on the other hand, it also creates an incentive to hoard the asset, limiting transactions.
        \item {\bf Unlimited  - Deterministic.} Very few (eventually not broadly adopted) digital currencies based on blockchain attempted to create unlimited supply, like Dogecoin or Freicoin.
        \item {\bf Pre-mined} Some altcoins (with the purpose of funding the development of the platform, or with the sole idea of profiting) have distributed all the assets before the starting of the system. Then, a reward system induces some kind of redistribution. 

\end{enumerate}

\section{Extensibility}\label{extens}
 The alignment of the interoperability, intraoperability, governance and script language determine the future ecosystem of
 the blockchain network and the integration possibilities of variety of blockchain related technology. Figure \ref{taxonomy-Matrix} illustrates the subcomponents forming the component Extensibility: 
\begin{itemize}
       \item[1] Interoperability
       \item[2] Intraoperability
       \item[3] Governance
       \item[4] Script Language
\end{itemize}

\sloppy \subsection{Interoperability}\label{inter}
Interoperability illustrates the overall capability of blockchains to exchange information with other systems, outside of blockchains.
It allows inflow, outflow and information retrieval of data providers that are not necessarily a blockchain-based system, e.g.~financial data providers\cite{dilley2016strong}.
We identify three possible layouts for \textit{Interoperability}:

\begin{enumerate}
        \item{\bf Implicit interoperability.}  It occurs when the smart contracts that specify conditions under which a particular transaction (or event) is to take place can be written in a  Turing-complete blockchain script language. In this context, implicitly any kind of condition can be specified, even those involving specific status in other systems. This implies an (albeit cumbersome) way of interaction from a blockchain solution to any API tool or interface. 
        \item {\bf Explicit interoperability.} If the script language is not Turing complete or the system has specific tools implemented that enable interoperability with the real world (like Bitcoin with Counterparty), then we talk about explicit interoperability, as it is brought purportedly into the system and one of its design principles.  
        \item {\bf No Interoperability.} A blockchain without any kind of possibility to interact with other systems. As implemented, Bitcoin in absence of external solutions (i.e. off the chain layers) has no interoperability implemented. It applies to most existing blockchain-based systems whose script language is not Turing complete.
\end{enumerate}

\subsection{Intraoperability}\label{intra}
Intraoperability illustrates the overall capability of blockchains to exchange information with other blockchains.
It allows inflow, outflow and exchange  of data between different blockchains\cite{chen2017inter}.
We identify three possible layouts for \textit{Intraoperability}:
\begin{enumerate}
        \item {\bf Implicit intraoperability} It occurs when the smart contracts that specify conditions under which a particular transaction (or event) is to take place can be written in a  Turing-complete blockchain script language. In this context, implicitly any kind of condition can be specified, even those involving specific status in other blockchains.
        \item {\bf Explicit intraoperability} If the script language is not Turing complete but is specifically designed to allow for intraoperability, then we talk about explicit intraoperability, because it is brought purportedly into the blockchain and it is  one of its design principles. An example of this is Bitcoin with Counterparty.
         \item {\bf No intraoperability} A blockchain without any kind of possibility to interact with other blockchains. As implemented, Bitcoin in absence of external solutions has no intraoperability implemented. Solutions for non intraoperable blockchains resort on: 1) Trusted proxies to connect blockchains; 2) Pegged blockchain systems; 3) Distinguishing tokens in the same blockchain based system. 
\end{enumerate}

\subsection{Governance}\label{governance}
Effective governance rules are crucial for the successful implementation of the blockchains and for their capability to adapt, change and interact. As the blockchain deployment structures (public chain, private chain, consortium chain) are different, their management patterns are also quite different.
We identify two type of governance rules: 1) {\it technical rules} of self-governance defined by the participants. Technical rules are composed of software, protocols, procedures, algorithms, supporting facilities and other technical elements; 2)  {\it regulatory rules} defined by external regulatory bodies composed of regulatory frameworks, provisions, industry policies and other components \cite{atzori2015blockchain,davidson2016disrupting,wright2015decentralized}. 
Regulatory rules are by definition not technical in nature and therefore outside the scope of this taxonomy. We focus instead on techical rules which are particularly interesting for their  feedback loop with the proposed technological solutions.
We identify three possible layouts \textit{Governance}:
\begin{itemize}
         \item {\bf Open-source Community.} In this case, open communities of developers (following open-source principles) and validators (very often in coordination with the blockchain foundation) coordinate upgrades and technical adjustments of the blockchain. For example, Bitcoin is mainly maintained by a team of core developers who in coordination with miners agree on changing parameters or other settings of the Bitcoin network. Also Ethereum and Hyperledger (backed up by the Linux Foundation) follow an open-source community model. 
         \item {\bf Technical}.  Since the blockchain technology is very versatile and can be applied to many business cases, enterprises with a strong technical strength (e.g., IBM and Microsoft) have proposed themselves as technical solution providers for blockchain architectures (proprietary hardware and software systems and basic services). In these cases, the technical rules of blockchain governance are dictated by the companies according to their business goals. For example, in 2015 Microsoft collaborated with ConsenSys to create the Ethereum blockchain technology service and took it as part of the Microsoft Azure service (EBaaS) to provide distributed ledger technology trials for enterprise customers, partners and developers. Moreover, in order to protect their proprietary blockchain architectures, these companies generally apply also for patents. According to \cite{WEF_Report} 2,500 patents on this topic have been filed from 2014 to 2016.
         \item {\bf Alliance}. This is the blockchain governance model proposed by industry consortia (e.g., B3i, R3) composed of companies with common business or technological progress demands. The alliance mode has the scope to sharing technology platforms to build common business models and standards. Only companies that meet certain criteria (e.g., payment of the fees, qualification of the organisation) are legitimised to collaborate to set technical rules of blockchain governance.  Those companies join together to promote commercial and technological progress in the area of blockchain under mutual benefit and common contribution.
\end{itemize}

\subsection{Script Language}\label{script}
Widespread programming languages are Turing-complete, which in formal terms refers to the fact that it is possible to implement an  algorithm on it to simulate any Turing machine. These are therefore general purpose languages, in which arbitrary computations can be performed. Languages that are not of this kind, are so because of design reasons which aim at prevent specific behaviours of code execution, like undefined termination.
  
Blockchain systems allow to modify the conditions under which certain information (e.g. transactions) will be included into the public record. These conditions must be specified in an algorithmic manner, and in some contexts are termed \textit{smart contracts}. 
These algorithms are elicited in a \textit{scripting language} designed purely for this purpose. 
Therefore the intended flexibility given to the users, with respect to the scope that the algorithm can develop, affects tremendously the degree of freedom to create conditions for some actions to occur (on the one hand) and the hypothetical computational effort that may be necessary to assess if a particular condition is fulfilled or not \cite{kim2017perspective}. 

It is worth remarking here that how limited is the scope of the scripting language is another design decision developers must carefully choose before the implementation of the blockchain, as abrupt changes (or bugs) may deride the logic of particular transactions.
We identify four possible layouts for \textit{Script Language}:
\begin{enumerate}
        \item {\bf Turing Complete.} Ethereum refers to a suite of protocols that define a platform for decentralised applications. With respect to scripting languages, on the one end of the spectrum, the Ethereum Virtual Machine (EVM) can execute code of arbitrary algorithmic complexity. In the terms described above, Ethereum is ``Turing complete'', because developers can create applications, which runs on the  EVM. Furthermore, Counterparty also uses Ethereum's entire smart contract platform to enable users to write Turing completeness for smart contracts.
        It has been pointed out that there exist scalability and security concerns regarding the usage of Turing-complete for scripting languages in blockchain systems \cite{atzei2017survey}. As of this writing, these have not been resolved.
 \item {\bf Generic Non-Turing Complete}. When designing Bitcoin, a decision was made to keep the scripting language limited in scope, to allow for a low impact of these calculations in the efficiency of the system. It is therefore a non-Turing complete language, and most blockchain implementations have followed this path. There is no connectivity in these to so-called ``oracles'' that allows obtaining data from sources that gather data which is exogenous to the blockchain.
 \item {\bf Application-specific Non-Turing Complete}. There are some non-Turing complete languages that are more expressive than the generic ones and purposely designed for certain cases. By restricting the language to be only able to write programs relevant to specific limited cases, the potential outputs of those programs becomes predictable. This allows those outputs to be queried and easily analysed. One example is Digital Asset Modelling Language (DAML) which is designed to codify only financial rights and obligations for execution in private networks. DAML is also more expressive than Bitcoin' script language and easier to read from a non techical audience.
       \sloppy \item {\bf Non-Turing Complete + External Data.} There exists a third category barely used so far that (while keeping the nature of the scripting language non-Turing complete) allows for existence of oracles. These oracles are considered trustful sources and add a layer of simplification on the validation to be performed by the language, empowering above Turing-completeness (as long as the oracles are reliable). This layout is then ``non-Turing Complete + External Data''.
\end{enumerate}



\section{Security and Privacy}\label{secpriv}
{The recent evolution and new implementations of blockchain systems bring risks, both technical and operational, associated with security and privacy. Thus, we group together security and privacy as two interrelated faces of the same problem.
Similarly, ISO TC 307 has created a dedicated Working Group on ``Security and Privacy'' \cite{iso.org}.}

{Security of blockchain systems is a matter of significant concern. Crypto currencies, the most widely deployed application of blockchain systems, have suffered from cyber attacks which became possible because of sensitive data mismanagement and the flawed design of the systems \cite{lin2017survey}. Without going into the detailed distinction between ``risks'', ``threats'', ``attack surfaces'' and ``vulnerabilities'', security of blockchain systems concerns:	1) Information mismanagement (alternation, deletition, distruction, disclosure etc.); 2)	Implementation vulnerabilities (including cryptomechanisms implementation vulnerabilities, run-time leakage of information etc. ); 3)	Cryptographic mechanisms mismanagement (including use of weak algorithms, key disclosure); 4)	User privileges mismanagement. 
For a recent comprehensive survey specifically targeting to the security and privacy aspects of Bitcoin and its related concepts we refer the readers to \cite{conti2017survey}.
With regard to privacy we refer to the
``freedom from intrusion into the private life or affairs of an individual when that intrusion results from undue or illegal gathering and use of data about that individual'' (ISO 25237 and other ISO standards). These privacy principles apply to any ICT system containing or processing PII, including blockchain systems.}
Figure \ref{taxonomy-Matrix} illustrates the subcomponents forming the component Security \& Privacy: 
\begin{itemize}
        \item[1] Data Encryption
        \item[2] Data Privacy
\end{itemize}

\subsection{Data Encryption}\label{dataencryp}
{By \textit{Data Encryption} we refer to cryptographic primitives. To ensure authenticity, integrity property and order of events, cryptographic primitive (cryptographic algorithms) are used.  For example, Bitcoin blockchain uses ECDSA digital signature scheme for authenticity and integrity, and SHA-2 hash function for integrity and order of event. Hash functions are also commonly used as a part of Proof-of-Work consensus mechanism.} 
We identify two major layouts for \textit{Data Encryption}:
\begin{enumerate}
            \item  {\bf SHA-2.} SHA stands for Secure Hash Algorithm. In its two incarnations,  SHA-256 and SHA-512, SHA (originally developed by the National Security Agency, USA) is the most widely variants for hashing functions \sloppy \cite{crosby2016blockchain, harvey2016cryptofinance} having first been used in Bitcoin. When issued to hash transactions, it requires a piece of information from the issuer, i.e.~the public key for the validation to take place\cite{meiklejohn2015privacy}. 
            \item {\bf ZK-SNARKS.} The Zero-Knowledge - Succinct Non-interactive Argument of Knowledge is a newer technology where no data whatsoever has to be provided to validate a specific hash \cite{ben2014scalable}. With the hashed message and the encrypted one, is sufficient as a proof to generate the validation. This anonymises  much more the individual information. 
\end{enumerate}

\quad

\subsection{Data Privacy}\label{dataprivacy}
 Although public/private key infrastructures and other measures like hashing functions should 
ensure that only the intended recipient can read the message and have access to the content of the transaction, the research shows that blockchain transactions (for e.g., in Bitcoin)  can be linked together in order to extract additional information and eventually also the identity of the participants \cite{tasca2016evolution}.  Indeed, there exists an inevitable tradeoff between a decentralised peer-validate system and the security and privacy of information. 
In this regard, several alternative solutions have been proposed to ``encrypt'' the data in such a way that even though computations and transactions occur in plain sight, the underlying information is completely kept obfuscated. Obfuscation is a way of turning any program into a ``black box''. This is equivalent to the original program: runs the same ``internal logic'' and provides the same outputs for the same inputs. But information on the data and processes is inaccessible. Of course there exists a strong interrelation between \textit{Data Privacy} and \textit{Data Encryption}.
According to the solutions proposes so far to enhance \textit{Data Privacy}, we identify two possible layouts:
\begin{enumerate}
             \sloppy \item {\bf Built-in data privacy}. With built-in data privacy we include all those blockchains that by default provide obfuscation of information. For example ZeroCash uses built-in zero-knowledge cryptography to encrypts the payment information in the transactions {\cite{sasson2014zerocash}}. Although ZeroCash payments are published on a public blockchain, sender, recipient, and amount of a transaction remain private. Alternatively, blockchains like Enigma (a project that seeks to implement the secret sharing DAO concept)\cite{wit2017dao} uses built-in secure multi-party computation guaranteed by a verifiable secret-sharing scheme. In this case, the data can be split among $N$ parties in such a way that  $M$ $<$ $N$ are needed to cooperate in order to either complete the computation or reveal any internal data in the program or the state. But $M$-$1$ parties cannot recover any information at all (which implies the need of trust on the majority of the participants to be honest). Finally, CORDA by R3 proposes a Node to Node ($N$-to-$N$) system characterised by encrypted transactions where only the parties involved in the transaction have access to the data { \cite{hearn2016corda}}. This is suitable for financial transactions 
where a high degree of confidentiality is required. Third parties like central banks or other market authorithies may have access to the data by invitation only.
            \item {\bf Add-on data privacy}. In this case, pseudonymous or public blockchains must resort on external solutions in order to obfuscate the information. One method is the \textit{mixing} service like Coinjoin. The principle behind this method is quite simple: several transactions are grouped together so to become a unique $M$-to-$N$ transaction. If for example, Alice wants to send one coin to Bob, and Carla wants to send one coin to David, a mixing transaction could be established whereby the addresses of Alice and Carla are both listed as inputs, and the addresses of Bob and David are listed as outputs in one unique transaction. Thus, when inspecting the $2$-to-$2$ transaction from outside it is impossible to discern who is the sender and who the recipient {\cite{Mixingservice} \cite{Coinjoin}}. Alternative to  the \textit{mixing} service, the \textit{secret sharing} allows data to be stored in a decentralised way across $N$ parties such that any $K$ parties can work together to reconstruct the data, but $K$-$1$ parties cannot recover any information at all. Alternative add-on data privacy tools are \textit{ring signatures} {\cite{noether2016ring}}
            and \textit{stealth addesses} 
            {\cite{moser2017anonymous}}
            which hide the recipient of a transaction and can be used by any blockchain. {Ring signatures - firstly introduced by \cite{rivest2001leak} -}        
            and its variant (linkable ring signatures) allow to hide transactions within a set of others' transactions. In this case the transaction is tied to multiple senders' private keys but only one of them is the initiator. Thus, the verifier may only identify that one of them was a signer, but not who exactly that was. In the case of stealth addresses, a receiver generates a new dedicated address and a ``secret key'' and then sends this address to someone who he wants payment from. The sender use the address generated by the receiver plus a ``nonce'' (one time random number) in order to generate the address he/she will send funds to. The sender communicates the nonce to the receiver wwho can unlock the address by using the nonce and the secret key generated earlier. {Monero (https://getmonero.org/) is an example of blockchain that aims to achieve privacy through the use of traceable ring signatures and stealth addresses}.
\end{enumerate}

\section{Codebase}\label{code}
The codebase of the blockchain technologies delivers information about which challenges a developer could face and what kind of changes the underlying programming language could undergo. Therefore the main component ‘Codebase’ is essential to align and increase the efficiency of blockchain related IT architectures. Figure \ref{taxonomy-Matrix} illustrates the subcomponents forming the component Codebase:
\begin{itemize}
        \item[1] Coding Language
        \item[2] Code License
        \item[3] Software Architecture
\end{itemize}



\subsection{Coding Language}\label{cod-lang}
Coding language illustrates the interconnectivity of programming languages of the blockchain technologies.
We identify two possible layouts for \textit{Coding Language}:
\begin{enumerate}
    \item {\bf Single Language.} Bitcoin has released The Bitcoin Core version 0.13.1 with the underlying coding language C++. As Bitcoin is open source, implementations occurred (much less popular than the original codebase) in different languages (like Java).  
     \item {\bf Multiple Languages.} Ethereum uses C++, Ethereum Virtual Machine Language and Go, which enables more interaction with other languages. Stellar maintains JavaScript, Java, and Go-based SDKs for communicating with Horizon. There are also community-maintained SDKs for Ruby, Python, and C-Sharp.
\end{enumerate}

\quad

\subsection{Code License}\label{license}
The Code License illustrates the possibility of changes to the source code of the underlying technology.
We identify thee possible layouts for \textit{Code License}:
\begin{enumerate}
        \item {\bf Open Source.} Regardless of the exact licence used for specific projects, we refer only to the openness in the source code as the only differentiating  factor. Bitcoin core developers have continuously  licenced the source code under the  MIT licence. Counter-intuitively, a permissive licence like the MIT  one (in which other developers can take the source code and fork it) eventually prevents multiple implementations. It also allows for continued development, larger code growth and allows adoption at a faster pace. Furthermore, Ripple and Stellar have licensed their codes with the ISC License. The ISC license is another permissive licence.  
        \item {\bf Closed Source.} For private implementations of blockchain-based systems, the source code is not necessarily openly distributed. {Just as an example, most the blockchains running  on the Ethereum Enterprise Alliance, rather than on the public Ethereum blockchain, use closed source codes.}
        In this case, risking the existence of unadressed bugs or unreported characteristics that may violate the expected conditions of use and functioning, the code may be kept outside of reach for users. 
\end{enumerate}


\subsection{Software Architecture}\label{Consensus-Engine}
The {\it Software Architecture}  refers to the high level structures of the blockchain system. Each structure comprises software elements, relations between them, and the properties that elements and relations give. The choice of the software architecture is very important in order to better manage changes once implemented. Software architecture choices include specific structural options among the  possibilities that are available for software design.

We identify two possible layouts for \textit{Software Architecture}:
\begin{enumerate}
            \item {\bf Monolithic Design}. In this case, all the aspects of a decentralised ledger (P2P connectivity, the ``mempool'' broadcasting of transactions, criterion for consensus on the most recent block, account balances, nature of smart contracts, user-level permissions, etc.) are  handled by a blockchain built as a single-tier software application without modularity.  \sloppy Examples of blockchains with monolithic design include Bitcoin and Ethereum. These architectures suffer from lack of extensibility on the long run. 
 \item {\bf Polylithic Design}. The Polylithic approach decouples the consensus engine and P2P layers from the details of the application state of the particular blockchain application.
For example, in Tendermint  the blockchain design is decomposed. It  offers a very simple API 
between the application process and its application-agnostic "consensus engine" (TenderminCore) which enables to run Byzantine fault tolerant applications, written in any programming language, not just the one the consensus engine is written in. Also Hyperledger {Fabric} follows a polylithic design as it is composed of interchangeable modules representing different components of blockchain technology.  
\end{enumerate}

\section{Identity Management}\label{identity-}
The main component \textit{Identity Management} ensures secure access to sensitive data to establish a suitable governance model for the blockchain. {This is a complex matter, as different levels of authority, accountability and responsibility are attached to different type of participants (e.g. users, administrators, developers, validators, etc). Generally, the set of rules are defined and enforced through mechanisms intrinsic to the system itself (on-chain governance)}.
{The subcomponents also eventually determine the concept of digital identity that users end up having within the systems}. Figure \ref{taxonomy-Matrix} illustrates the subcomponents forming the component Identity Management: 
\begin{itemize}
        \item[1] Access and Control Layer
        \item[2] Identity Layer
\end{itemize}

\subsection{Access and Control Layer}\label{access-control}

When establishing the right governance structure for a blockchain it is important to consider the ledger construct.  Depending on its purpose, the ledger could be run by a central authority and governed by it or it could be run in a decentralised fashion according to a set of governance rules adhered to and enforced by participants on the blockchain network. The governance structure determines
the authorisation and the control policy management functions. Those rules provide permission for users to access to or use blockchain resources. Those are a set of rules that manage user, system and node permissions that must be followed in security-related activities. 
Blockchains may have different permissions according to which access and control to data is allowed. The distinguishing features must answer to the following questions:
\begin{itemize}
\item Which users have ``read" access? 
\item Which users have ``write" access?
\item Is it there anyone who can ``manage consensus" (i.e., update and maintain the integrity of the ledger)?
\end{itemize}
According to the set of governance  rules, we may have different system designs that reply to the above questions in a different way in order to better serve either a public or a private interest of either a \textit{general} 
(like in the case of Ethereum) or a \textit{special}  (like in the case of Corda) purpose.
On one side, private blockchains are generally those with a set of constrained ``read/write" access alongside a consensus algorithm which allows only a pre-selected group of people to contribute and maintain the blockchain integrity. Instead, public blockchains do not control  ``read/write" access or in the consensus algorithm for any given set of participants. Nevertheless, this does not mean that certain permission structures can not be implemented as part of a specific application.


Although different variations are possible, the authority to perform transactions on a blockchain generally  belongs to one of the following main models of the \textit{Access and control Layer}\cite{guegan2017public}:
\begin{enumerate}
        \item {\bf Public blockchain}. In this case, there is no preference in  access or in managing consensus. All participants (nodes), 
         have "read/write" access and without any control can contribute to the update and management of the ledger. An example of blockchain in this area is Bitcoin where every participant can either choose just to use the blockchain to exchange Bitcoins (or other data on the top of it, in general by means of third-party technologies), run a full node or even become a miner to participate in the process of transaction validation.
         \item {\bf Permissioned Public blockchain}. In this case ``read" access is enabled for all users, however ``write'' access and/or ``consensus management" require permission by a pre-selected set of nodes. Ripple belongs to this group as to validate transactions a participant need to be part of the so-called \textit{Unique Node List}. Some other examples include Ethereum and Hyperledger Fabric which is used for the exchange of tangible (real estate and hardware) with intangible (contracts and intellectual property) assets between enterprises. 
 \item {\bf Permissioned Private blockchain}. 
        In this case, "read/write" and "consensus management" rights can only be granted by a centralised organisation.  
         An example is Monax (formerly known as  Eris).
\end{enumerate}


\subsection{Identity Layer}\label{identity}
The onboarding and offboarding of nodes / entities to the blockchain networks is handled differently by the various software solutions. 
{By identification we mean the capability to identify an entity uniquely in a given context. Digital identity can be defined as a set of identifying attributes for an entity that together enable the unique identification of the entity in a context (UID).  A vital part of any identity system (and most information systems) is that a UID is managed throughout the entity’s lifecycle to protect it from negligence and fraud, and to preserve the UID’s uniqueness.  A UID can then be assigned to the identity and used to link or bind the entity to the claimed identity and to any digital credential (software or hardware) issued to the entity. This digital credential acts a trusted proxy for the physical or logical entity and is used to support a wide range of personal and trust-related functions such as authentication, encryption, digital signatures, application logins and physical access control.}

AML and KYC procedures -- 
generally required to proceed personal related data e.g. medical data, bank information or other personal related data --, are the key aspects to consider when looking at the \textit{Identity Layer}.
We identify two possible layouts:
\begin{enumerate}
        \item {\bf KYC/AML.} 
        {Compliant blockchains have the ability to validate organisations and their attribute data from authoritative sources to ensure the quality of data written to the blockchain and linked to identifiers in the blockchain. An example is} Stellar that sets requirements for all integrators to implement Know-Your-Customer (KYC)/Anti-Money-Laundering (AML) identity verification process to increase the transparency of the stellar network participants. Furthermore, Ripple forces its financial services partners to implement an identity layer to verify the user information.The financial services partners have to do a due diligence, depending on the requirements they must fulfil.
        \item {\bf Anonymous.} {In general the common misunderstanding of the anonymity level within Bitcoin networks is that the majority of the users do not distinguish between anonymity and the pseudo-anonymity. In light of this, the Bitcoin protocol has no identity layer to identify the users. Those circumstances could benefit misuse of Bitcoins and money laundering activities through this blockchain network, but to control approaches to anonymity in Bitcoin and other cryptocurrencies\cite{maurer2016survey}. Regal Reid and Martin Harrigan \cite{reid2013analysis} have been able to demonstrate that several pseudonymous
        addresses can be linked to one single user. See also \cite{tasca2016evolution} for a Bitcoin transaction-path driven users identification method.}
\end{enumerate}


\section{Charging and Rewarding System}\label{reward}
Blockchain systems incur in operational and maintenance costs that are generally absorbed by the participants to the network. Different kind of cost models are applied according to: 1) the architectural configuration design; 2) the governance system; 3) the data structure and the computation required on-chain. One of the 
 cost items which is common to the wide majority of the blockchains is the verification cost. This is required to sustain the validation process of the transactions that compete to be appended (and never removed) to the ledger.
The potential financial costs incurred when taking part of a blockchain platform, require an incentive scheme that maintains consistency of the cost structure across the different stakeholders. Figure \ref{taxonomy-Matrix}  illustrates the subcomponents forming the component Charging and Rewarding System:
\begin{itemize}
   \item[1] Reward System
   \item[2] Fee System
   \begin{itemize}
       \item[2.1] Fee Reward
       \item[2.2] Fee Structure
       \end{itemize}
\end{itemize}

\subsection{Reward System}
 This subcomponent illustrates the rewarding mechanisms automatically  put in place and triggered  by the systems in order to compensate active members contributing to data storage or transaction validation and verification.
We identify two possible layouts for \textit{Reward System}:
\begin{enumerate}
    \item {\bf Lump-sum Reward.}  Individuals taking part of the storage, validation or verification process (e.g., in the Bitcoin verification is only  rewarded to users called \textit{miners}) may be rewarded for their action. For example, in Bitcoin, the first transaction in each block is called \textit{coinbase}, and the recipient is the user (or users) who created the block, that in this regard is a set of transactions verified by said user(s).
    The lump-sum reward can be fixed like in Enigma or variable like in Bitcoin. 
    \item {\bf Block + Security Reward.} In other  \sloppy blockchain-based technologies, like Ethereum, the blockchain rewarding system includes, besides the block reward, a reward for including in the validation forked blocks that are still valid. The design idea is to incentivise cross-validation of transactions (crucial in a setting where validation can be arbitrarily costly)\cite{timmerman2017ethereum}.
\end{enumerate}

\quad

\subsection{Fee System}\label{fee-system}
Other kind of rewards are those provided directly by the users to other participants of the system when launching any request in the network for storage, data retrieval, or computation and validation. With regards to this, we identify two sub-subcomponents: \textit{Fees Reward} and \textit{Fee Structure}.

\subsubsection{Fee reward}\label{fee-reward}
\textit{Fees Reward} describes the nature of the fees that the users are required to contribute when using a blockchain. {The fees system has been shown to play an important role in the way verifiers do  \cite{moser2015trends}} and may \cite{carlsten2016impact} behave. This kind of design-time consideration ought not to be neglected, as it is usually the case.
We identify three possible layouts for \textit{Fees Reward}:
\begin{enumerate}
    \item {\bf Optional Fees. } In Bitcoin and related technologies \cite{FeeMining} users can optionally pay a voluntary fee for the validation process. This fee is optional, but it is assumed that the larger the fee is, the lower is the processing time it will take to be added to a block, as miners will be more incentivised to do so. Moreover, given that the coinbase reward halves approximately every four years, currently, the reference Bitcoin client refuses to relay transactions with zero or no fees.
    \item {\bf Mandatory Fees.} Some systems like Stellar force all users to include fees in any transaction added into the system.  
    \item {\bf No Fees.} In comparison, the Hyperledger Fabric is a blockchain solution for businesses, which combines a permissioned network and an identity layer without any transaction fees. 
\end{enumerate}

\subsubsection{Fee structure}\label{fee-structure}
When provided by the system, fees can follow either a fixed or a variable structure.
There are two alternative layouts for \textit{Fees Structure}:
\begin{enumerate}
    \item {\bf Variable Fees.} In this case, the fee is somehow linked to the "size" of the request. In Bitcoin, the larger the transaction size, the higher will be the fee the user shall pay in order to compensate for taking up space inside the block. Miners usually include transactions with the highest fee/byte first. The user can decide how many Satoshis (0.00000001 Bitcoins) wants to pay per byte of transaction. For example, if the transaction is 1,000 bytes and the user pay a fee of 300,000 Satoshis, he/she will be in the 300 Satoshi/bytes section (300,000/1,000=300.00). At the time of writing, this implies that the transaction will be included in the next 2 block transactions (i.e., within 20 minutes).
    However, to avoid queuing, the user can increase the fee. The fastest and cheapest transaction fee is currently 360 Satoshis/byte. For an average transaction size of 226 bytes, a fee of 81,360 Satoshis is currently the cheaper fee in order to get the transaction included in the first available block without delays. Also other blockchains apply variable fees and follows similar rules as Bitcoin. 
    \item {\bf Fixed Fees.} In this case, the fee is linked to the request, not to its "size". For example, in Enigma every request in the network for storage, data retrieval, or computation has a fixed price, similar to the concept of Gas in Ethereum. However, since Enigma is a Turing-complete system, the fee can be different depending on the specific request. Another example of blockchain with a fixed transaction fee is Peercoin which required a fixed 0.01 PPC per kilobyte.
\end{enumerate}


\section{Conclusion}

In the 21st century, the blockchain technologies will athwart affect all business areas: financial services \cite{alvseike2017blockchain,scott2016can,evans2015bitcoin,quintana2014merger}, IoT \sloppy \cite{boudguiga2017towards,dorri2017towards}, consumer electronics \cite{andrews2017utilising}, insurances \cite{mainelli2014chain, stellnbergerinsurance}, energy industry, logistics \cite{badzar2016blockchain,hackius2017blockchain}, transportation, media \cite{kotobi2017blockchain}, communications \cite{plant2017implications}, \cite{antorweep}, entertainment, healthcare \cite{kuo2017blockchain}, automation, and robotics will be involved. After the advent of Internet, it currently represents the most prominent technology and it will shape the upcoming products and services in every industry field. 
Since the introduction of Bitcoin in 2009, the awareness of blockchain technologies has considerably increased.
During the initial phase, the first mover regarding the adoption of blockchain was the financial industry. This is explained by the fact that blockchain enables cost reduction and increases the efficiency in several business processes (both internal and external) for financial institutions. 
An example of the big impact of blockchain in the financial industry regard the networks of global payments which involve money transactions in exchange of goods, services or legal obligations between both individuals or economic entities. Beyond payments, blockchain allows real-time settlements, which reduces operational costs for the banks. Furthermore, the immutability of the blockchain reduces the risk of fraud, as a consequence banks can use  sophisticated smart contracts to capture digital obligations and to eliminate operational errors.  Global payments are just a fraction of the overall use cases in the financial industry. Moreover, many other industries, including the public sector, are now looking at blockchain-enabled solutions for their own processes. 
This tremendous trend is causing a proliferation of multiple blockchain {architectures} which often are not interoperable and are built  according to different engineering designs. Lately, software architectures, companies and regulators realised the need for standardisation of some of their components. 
This is becoming a necessary step for the blockchain in order to: 1) gain global adoption and compatibility, 2) create cross-industry solutions, 3) provide cost-effective solutions. As always with standardisation processes, their creation must be a necessary equilibrium between different parties. But in this particular case, the open source community that brought forward this disruptive technology and which continuously develops most implementations should play a crucial role, as heralds of the advancement of blockchain.

{Based on the review of the current literature on blockchain technologies, our work is an early stage analysis across existing software architectures  with the aim to propose a taxonomy: a reference architectural model for blockchains and their possible configurations.  Based on component-based design, the blockchain taxonomy decomposes the blockchains into individual functional or logical components and identifies any possible different layout. The blockchain taxonomy proposes to assist in the exploration of design domains, in the implementation, deployment and performance measurement of different blockchain architectures. Figure \ref{taxonomy-Matrix} illustrates the blockchain taxonomy tree resulting from our analysis.} 

 {Our work sheds light on the current proliferation of non-interoperable blockchain platforms and on the need (for) and current discussions (about) blockchain standards. Although our work contributes to the ongoing efforts on setting blockchain standards, we do not conclude by saying that we need a set of standards \textit{now}. This process generally takes several years in order to produce concrete solutions (even 10 years for complex subjects). Therefore, we think that our taxonomy represents a timely honest intellectual exercise to be used as preliminary supporting material for all those interested in reducing blockchain complexity. At the same time,  we are aware that our taxonomy tree, although hopefully very useful, is very preliminary and likely the first version of subsequent more complex evolutions.}

\paragraph{Acknowledgements:} PT and CJT thank Harvey R. Campbell for his 
invaluable comments on a previous version of this paper. They also thank Alessandro Recchia and Thayabaran Thanabalasingham for their support and contribution.
PT acknowledges the University College London for financial support through the EPSRC program  EP/P031730/1.
CJT acknowledges the University of Zurich for financial support through the University Research Priority Programme on Social Networks.



\bibliography{general2,general3}

\begin{appendix}

\section{Blockchains Analysed for the Taxonomy}

\begin{table}[H]
\begin{center} 
\scalebox{0.7}{ 
\begin{tabular}{|>{\columncolor{White}}m{5cm}|>{\columncolor{White}}m{15cm}|}
\rowcolor{orange}
\hline
 \textbf{\LARGE{Technology}}  & \textbf{\LARGE{Description}} \\\hline
	Bitcoin & Forerunner and by far the most widely used cryptocurrency.\\
\hline
 Dash & Privacy-centric digital currency. The digital currency Dash has different functionalities e.g. instant transactions. Dash is based on the Bitcoin source code, but it allows  anonymity while performing transactions.\\
\hline 
 Monero & Monero is an open-source digital currency, with the focus on decentralisation, scalability and privacy. It is based on the CryptoNote protocol, which has an enabled anonymous layer.\\
\hline 
 LiteCoin & LiteCoin is a P2P cryptocurrency and open source software project. The Litecoin is technically nearly identical to Bitcoin, except for the proof-of-work cryptographic function used.\\
\hline 
 Zcash & Zcash is a decentralised and open-source cryptocurrency, which combines privacy with selective transparency of transactions.\\
\hline 
 Peercoin & Cost-effective and sustainable cryptocurrency based on proof-of-stake.\\
\hline
	ColorCoin &	A concept that allows attaching metadata to Bitcoin transactions and leveraging the Bitcoin infrastructure for issuing and trading immutable digital assets that can represent real world value.\\
	\hline
	 Omnilayer (MasterCoin) &	A meta-protocol layer that enables new digital currencies, digital assets, and communication protocol to existing on top of the Bitcoin blockchain. \\
	\hline
	 NameCoin	& NameCoin is an asset registry on top of the Bitcoin Blockchain, which enables a decentralised domain name system.\\
\hline
 Counterparty &	Counterparty enables anyone to write specific digital agreements or programs known as smart contracts, and execute them on the Bitcoin blockchain.\\
\hline
	NXT (Ardor)	& NXT is a safe, transparent and decentralised system for sharing data and allowing payments to people all over the world. Ardor platforms enable smart contract functions with NXT.\\
	\hline
 Ethereum &	Ethereum is an open-source platform to build blockchain-based applications in different business fields.\\
	\hline
 Monax & Monax is an open platform for developers and DevOps to build, ship, and run blockchain-based applications for business ecosystems. Monax is known as a private blockchain offered by the monax company.\\
	\hline
 Cosmos & Cosmos is an architecture for cross-chain interoperability where independent blockchains can interact via an inter-blockchain communication (IBC) protocol, a kind of virtual UDP or TCP for blockchains
each driven by the Byzantine fault tolerant (BFT) consensus algorithm, similar to Tendermint.\\
\hline
 COMIT & COMIT is a cryptographically-secure off-chain multi-asset instant transaction network (COMIT) that can connect and exchange any asset on any blockchain to any other blockchain using a COMIT cross-chain routing protocol (CRP).\\
\hline 
 Synerio & Synerio is intended to become a decentralised social network.\\
\hline
Ripple	& Ripple is a global real-time financial settlement provider.\\
	\hline
 Stellar &	Stellar is a decentralised multicurrency-exchange platform for people without access to the banking system.\\
	\hline
Hyperledger	& Hyperledger (or the Hyperledger project) is an open source blockchain platform, started in December 2015 by the Linux Foundation, to support blockchain-based distributed ledgers. It is focused on ledgers designed to support global business transactions, including major technological, financial, and supply chain companies, with the goal of improving many aspects of performance and reliability.\\ 
\hline
 Tendermint & Tendermint is a high-performance blockchain consensus engine that enables to run Byzantine fault tolerant applications written in any programming language. Tendermint  
is a partially synchronous BFT consensus protocol derived from the DLS consensus algorithm.\\
\hline 
 Corda & Open-source distributed ledger platform.\\
\hline{2-3} 
 Enigma & Distributed ledger technology, created by the MIT digital currency lab.\\
\hline
\hline
\end{tabular}}
\caption{\footnotesize{Blockchain Technologies.}} \label{DLT-techs}
\end{center}
\end{table}

\end{appendix}


\end{document}